\documentclass[twocolumn,showpacs,preprintnumbers,amsmath,amssymb]{revtex4}

\usepackage{graphicx}

\begin{document}

\title{Ferromagnetic Mn doped InSb studied at the atomic scale by cross-sectional STM}

\author{S.J.C. Mauger}
\author{J. Bocquel}
\author{P. M. Koenraad}
\email[]{p.m.koenraad@tue.nl}
\affiliation{Department of Applied Physics, Eindhoven University of Technology, Den Dolech $2$, $5612$ AZ Eindhoven, Netherlands}
\author{C. Feeser}
\author{N. Parashar}
\author{B.W. Wessels}
\affiliation{Department of Materials Science and Engineering and Materials Research Center,
Northwestern University, Evanston, Illinois 60208, USA}

\date{\today}

\begin{abstract}

We present an atomically resolved study of metal-organic vapor epitaxy grown Mn doped InSb that is ferromagnetic at room-temperature. Both topographic and spectroscopic measurements have been performed by cross-sectional scanning tunneling microscopy. The measurements show a perfect crystal structure without any precipitates and reveal that Mn acts as a shallow acceptor. The Mn concentration obtained from the cross-sectional STM data compares well with the intended doping concentration. No second phase material or (nano)clustering of the Mn was observed. While the pair correlation function of the Mn atoms showed that their local distribution is uncorrelated beyond the STM resolution for observing individual dopants, disorder in the Mn ion location is clearly noted. We discuss the implications of the observed disorder for a number of suggested explanations of the room-temperature ferromagnetism in Mn doped InSb grown by metal-organic vapor epitaxy. 

\end{abstract}


\maketitle

Dilute magnetic semiconductors (DMS) have attracted a strong scientific interest in recent years. Such materials can combine electrical, optical and magnetic properties that can be applied, for instance, in information processing.\cite{Ohno1998, Ohno1999} On the road to achieve room temperature applications, the highest Curie temperature (\mbox{$T_{C}$}) for DMS have been predicted for transition metal dopants in wide gap semiconductors. However transition metal dopants in wide gap semiconductors form deep levels in the band gap. Additionally, the formation of defects in the host material, such as for example As antisites and Mn interstitials in (Ga,Mn)As reduce the carrier concentration and lower the carrier mediated ferromagnetism.\cite{Dietl2000}
Interestingly, high Curie temperatures have been achieved for narrow gap (III,Mn)V semiconductors grown by metal-organic vapor epitaxy (MOVPE) with respectively \mbox{$T = 330$ K} for (In,Mn)As \cite{Blattner2002,Blattner2001} and more than \mbox{$T = 400$ K} for (In,Mn)Sb \cite{Parashar2010}, despite the prediction of a low \mbox{$T_{C}$} by mean field theory for DMS.\cite{Dietl2000} 
The position of the Mn acceptor level in narrow gap III-V semiconductors is predicted to be either shallow or resonant in the valence band and narrow gap III-V semiconductors have the advantage to remain highly conductive even for a high Mn concentration and therefore should allow for interesting magnetic properties.\cite{Wessels2008}

\begin{figure}[t]
\begin{center}
\includegraphics[width=82.5mm]{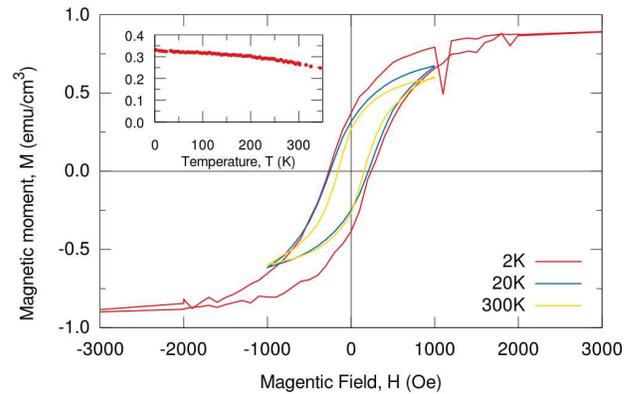}
\caption{\mbox{M(H)} hysteresis loops at \mbox{$T = 2$ K, $20$ K and $330$ K} of a Mn:InSb sample with a similar doping level as the sample measured with X-STM. The magnetic field is aligned along the [001] direction. The inset shows the temperature dependent remnant magnetization after a 100 Oe field has been applied.}
\label{SQUID curves}
\end{center}
\end{figure} 

In this paper we present an atomically resolved study of (In,Mn)Sb by cross-sectional Scanning Tunneling Microscopy (X-STM) \cite{Yakunin2004,Yakunin2005} to study the structural and electronic properties of this room-temperature ferromagnetic material. The electronic state and related incorporation site of the magnetic Mn atoms, the formation of clusters and second phases as well as the formation of additional defects are all important for a full understanding of the (magnetic) properties of (In,Mn)Sb. X-STM is an excellent technique to explore such properties. It is also interesting to study with X-STM the distribution of the magnetic atoms in the material. For instance, it was shown that percolation can play an important role in the Curie temperature of ferromagnetic materials.\cite{Sato2004,Kennett2004} Furthermore theoretical calculations have shown that disorder in the distribution of the magnetic dopant atoms in dilute magnetic semiconductors such as atomic scale clusters of pairs can lead to an increase of the Curie temperature.\cite{Bouzerar2004} Such atomic scale clustering has been suggested to be involved in the ferromagnetism observed in (In,Mn)As.\cite{Soo2004} In another approach that deals with a possible spatial disorder in the distribution of Mn atoms is proposed by Kennett et al.\cite{Kennett2004} They developed a two-component-model whereby disorder in the Mn ion locations can lead to spatially inhomogeneous local magnetizations that are strongly correlated with the local charge density. 
Thus the incorporation of transition metal dopant atoms and their distribution can play a central role in determining the magnetic properties of the DMS. The question arises as to the nature of the dopant disorder at concentrations of the order of \mbox{$\sim10^{20}$ cm$^{-3}$}. Cross-sectional scanning tunneling microscopy (X-STM) is well-suited to reveal the dopant distribution at high resolution.\cite{Yakunin2005,Kitchen2006,Richardella2010, Bozkurt2010} To the knowledge of the authors no spatial study at the atomic scale of Mn:InSb has been reported. Here we show that Mn is incorporated on the group III site and acts as a shallow acceptor. No indications were found for clustering, additional defects or second phases. Interestingly we found that the dopant distribution is uncorrelated for pair distances beyond \mbox{$3$ nm}. However local disorder, as is to be expected for an uncorrelated pair distribution function, is observed in the X-STM images of ionized Mn acceptors in (In,Mn)Sb. We will discuss the implications of the X-STM observations for a number of suggested explanations of the observed room-temperature ferromagnetism in MOVPE grown InSb.

The InSb films doped with Mn were grown by MOVPE on an undoped InSb substrate at a temperature of \mbox{$T = 400$ $^\circ$C}. The film thickness of our samples is \mbox{$500$ nm} and the Mn concentration measured by Energy Dispersive X-ray Spectroscopy (EDX) \mbox{$ 9 \times 10^{19}$ cm$^{-3}$}. Please note that the uncertainty in this value can be rather high because the data is taken close to the detection limit of this technique. Temperature dependent resistivity measurements show that the samples exhibit a metallic-like conduction which is consistent with the Mott limit indicating a critical hole composition of \mbox{$2.4\times10^{17}$ cm$^{-3}$}. \mbox{Fig. \ref{SQUID curves}} displays SQUID magnetometry at different temperatures that has been performed on a sample with a similar doping level as in the sample studied by X-STM. A clear magnetic hysteresis at room-temperature is observed. It is difficult to give an accurate estimation of the magnetic moment but a value of about 0.5\,$\mu_\textnormal{B}$ is obtained when assuming a homogeneous distribution of Mn atoms and relying on the EDX concentration. The magnetization remained after we used a number etching steps in order to remove possible second phase material on the surface. Moreover the remnant magnetization as a function of the temperature shows no signs of a second phase. Second phase material was also excluded by X-ray Absorption Spectroscopy (XAS) and X-ray Magnetic Circulair Dichroism (XMCD) measurements \cite{Parashar2009} in similar material grown in the same MOVPE system. The magnetization of the free carriers in the MOVPE material was furthermore corroborated by the presence of an anomalous Hall effect.\cite{Parashar2010} 

\begin{figure}[t]
\begin{center}
\includegraphics[width=82.5mm]{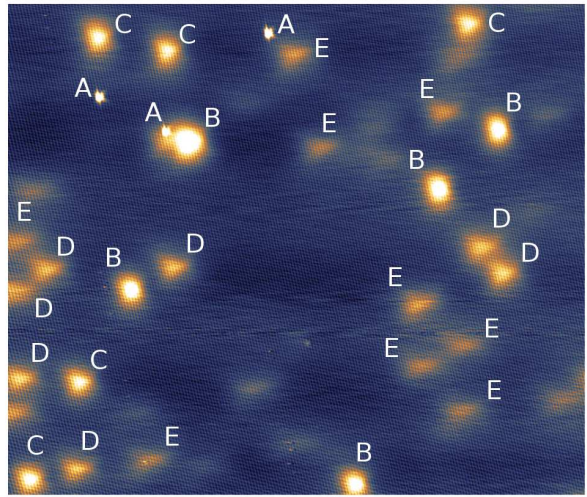}
\caption{Constant current X-STM topographic image (\mbox{$100$ nm$\times85$ nm}) of the (110) Mn:InSb surface. The indices correspond with the depth of the Mn acceptor below the (110) surface (\mbox{$V_{sample}$ = +$0.5$ V} and \mbox{I = $50$ pA}).}
\label{XSTM image}
\end{center}
\end{figure} 

For the X-STM measurements, the samples were cleaved in situ along their natural (110) cleavage plane under ultra high vacuum conditions (\mbox{$P$ $ <4\times10^{-11}$ mbar}) and the measurements were performed at 77 K using a commercial Omicron scanning tunneling microscope. In a very large number of scans taken at different positions in the epilayer and the substrate we do always find a perfect single zinc-blende phase with no evidence for a second phase or any (nano)clustering. A typical large scale atomically resolved X-STM image of the (In,Mn)Sb (110) cleavage surface obtained at constant current and positive sample voltage (empty state imaging) is shown in \mbox{Fig. \ref{XSTM image}}. The bright features correspond with substitutional Mn acceptors at different depths below the cleavage surface.  In previous X-STM measurements on (Mn,Ga)As we were able to prove that Mn acts as an deep acceptor that incorporates on a group III site as a Mn$^{2+}$ ion with a 3d$^{5}$ ground state.\cite{Yakunin2004} Depending on the polarity of the applied voltage, either the neutral or ionized charge state of the Mn acceptor is imaged. This is due to tip-induced band bending which can pull the Mn acceptor level below the Fermi level at a negative voltage bringing the Mn acceptor in a ionized state, whereas at positive sample voltage, the acceptor level of the Mn dopant can be pulled above the Fermi level bringing the acceptor in a neutral state. 
In \mbox{Fig. \ref{XSTM image}}, which is taken at positive sample voltage, we observe the Mn acceptors in their neutral state and the STM contrast of a single Mn reflects the charge distribution of the hole bound to the Mn acceptor. This is in excellent agreement with previously obtained XAS spectra that indicated the presence of Mn$^{2+}$ ions with a 3d$^{5}$ ground state, corresponding with an acceptor-like behavior, in similar MOVPE grown InSb samples. Previous STM studies on various acceptors with a different binding energy in III-V semiconductors have shown that the charge distribution of the hole bound to the acceptor is related to the binding energy of the acceptor.\cite{Celebi2010,Yakunin2004} At a fixed depth, the charge distribution of a deep acceptor (binding energy of the order of \mbox{$100$ meV} or more) is characterized by a bow-tie shape \cite{Celebi2010,Yakunin2004} whereas the shape of the charge distribution of a shallow acceptor such as Zn in GaAs\cite{Zheng1994} or Mn in InAs\cite{Marczinowski2008} is triangular. Our X-STM measurements clearly show a triangular contrast for the Mn dopants corresponding with an shallow acceptor state. This agrees with Obukhov et al.\cite{Obukhov1991} where it is shown that Mn in InSb acts as a shallow acceptor with a binding energy of about 9 meV.\cite{Obukhov2005,Teubert2009} If we image the cleaved surface at negative voltages (filled state imaging) we observe circular symmetric contrast. For Mn in GaAs we know that the acceptor at these tunnel conditions is ionized due to the band bending.\cite{Yakunin2004,Teichmann2008} The circular symmetric contrast is thus due to the influence of the Coulomb potential of the ionized Mn acceptor on the tunneling process.

\begin{figure}[t]
\begin{center}
\includegraphics[width=82.5mm]{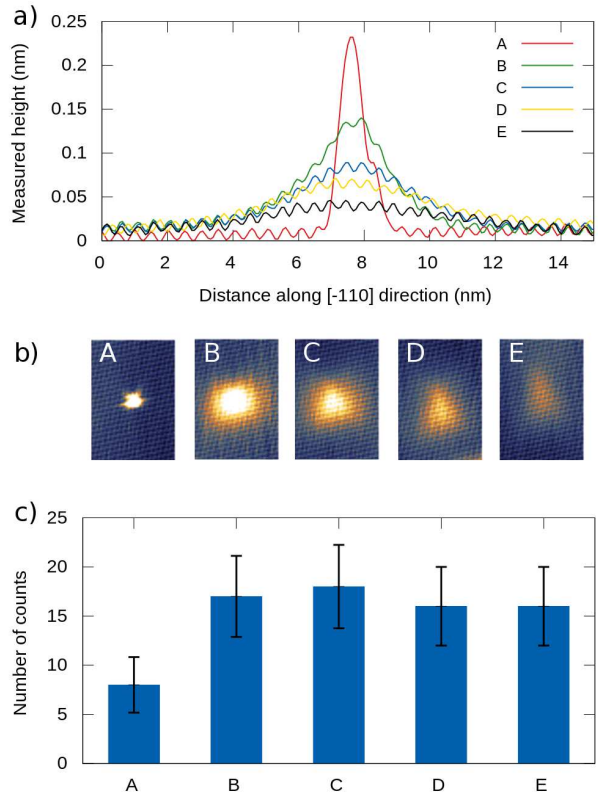}
\caption{(a) Topographic profile lines along the [110] direction for the various classes of Mn acceptors in InSb observed in \mbox{Fig. \ref{XSTM image}} which are all in their neutral charge state configuration. (b) Constant current STM images (\mbox{$10$ nm$\times7$ nm}) of the same classes of Mn acceptors. (c) Histogram of the number of Mn acceptors of each class observed in \mbox{Fig. \ref{XSTM image}}.}
\label{XSTM dopants}
\end{center}
\end{figure}

The differences in the shape and intensity of the Mn contrast observed in \mbox{Fig. \ref{XSTM image}} is related to the depth of Mn atoms below the (110) surface.\cite{Garleff2008} \mbox{Fig. \ref{XSTM dopants}(b)} shows the different classes of contrast of Mn acceptors that can be found in \mbox{Fig. \ref{XSTM image}}. The highly localized feature of class A in \mbox{Fig. \ref{XSTM dopants}(b)} corresponds to a Mn acceptor localized in the uppermost monolayer of InSb. The contrasts B-E all have an almost triangular shape but the intensity of the contrast varies strongly as shown by the profile lines taken in the 110 direction and shown in \mbox{Fig. \ref{XSTM dopants}(a)}. 
The Mn impurities with a less intense contrast are located deeper below the (110) cleavage plane than the impurities with a more intense contrast. Thus we conclude that the acceptors of class B-C-D-E are located respectively in the 2$^{nd}$, 3$^{rd}$, 4$^{th}$ and 5$^{th}$ monolayer below the (110) surface. Because Mn acceptors deeper below the cleavage surface are not resolved as clearly and distinctively as those in layer 1 to 5 they are not categorized. \mbox{Fig. \ref{XSTM dopants}(c)} shows the frequency distribution of the different classes of Mn contrast. 
The frequency of appearance is homogeneous for every class of Mn contrast except for the Mn in the uppermost layer which appears less than the other classes. The frequency of the Mn acceptors in the top five layers was used to determine the Mn doping concentration. The estimated Mn concentration based on the X-STM measurement shown in \mbox{Fig. \ref{XSTM image}} is \mbox{$3 \times10^{19}$ cm$^{3}$} which is somewhat lower than obtained form EDX measurements \mbox{$9 \times10^{19}$ cm$^{3}$} on this sample. This difference is very likely due to the uncertainties in: 1) the EDX measurements where the data is obtained close to its detection limit, 2) the uncertainty in the counting based on the X-STM measurements and 3) a gradient in the Mn concentration that we observed in this sample. Because we cleave the sample along the growth direction we get a cross-section through the grown epilayers and the substrate and thus various regions of the epilayer and the substrate have been scanned. While scanning the (110) plane in the [001] direction from the epilayer towards the substrate no clear (In,Mn)Sb/InSb interface was found. Instead we observe a monotonous decrease of the Mn concentration extending from the epilayer into the substrate. Mn atoms have been found at a distance of up to \mbox{$1500$ nm} from the top of the film. Because the Mn:InSb epilayer thickness is \mbox{$500$ nm} Mn atoms must have diffused by \mbox{$1000$ nm} into the substrate. The small difference between the melting point of InSb (\mbox{$527$ $^\circ$C}) and the growth temperature (\mbox{$400$ $^\circ$C}) might be related to this long range diffusion. 

\begin{figure}[t]
\begin{center}
\includegraphics[width=82.5mm]{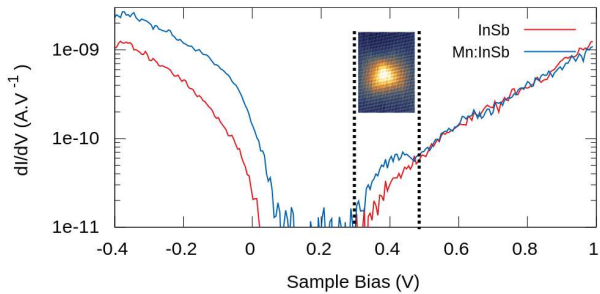}
\caption{Local tunneling spectroscopy spectra measured on Mn:InSb obtained with the set point at \mbox{$V_{sample}$ = $+1$ V} and \mbox{I = $200$  pA}. The red line represents the \mbox{$dI/dV$} spectra on the free InSb (110) surface whereas the blue line represents the \mbox{$dI/dV$} spectra taken on a Mn acceptor. The inset shows an empty state X-STM image of a Mn acceptor in InSb obtained at \mbox{$V_{sample}$ = +$0.5$ V} and \mbox{I = $50$ pA}.}
\label{XSTM spec}
\end{center}
\end{figure}

The sample was also analyzed by X-STM spectroscopy. In these measurements the feedback loop of the piezo-scanner is deactivated and at each point in the studied area the current is measured while the voltage is scanned. \mbox{Fig. \ref{XSTM spec}} shows the \mbox{$dI/dV$} spectraÕs taken on a free InSb (110) surface (red line) and on a Mn impurity (blue line). The current at negative sample voltage is due to the electrons tunneling from the filled states of the sample (i.e. valence band) to the tip whereas at positive sample voltage, the current is due to the electrons tunneling from the tip into the empty states (i.e. conduction band). 
The band gap of InSb, at the measurement temperature of \mbox{$77$ K}, can be determined from the range of low tunnel current and is about \mbox{$0.3$ eV}. This is a good agreement with the value of \mbox{$0.23$ eV} at \mbox{$80$ K} that is reported in the literature. 
%Although during the spectroscopy measurements the set point is chosen to keep the tip-sample distance constant as much as possible during the STS scan (\mbox{$V_{sample}$ = $+1$ V} and \mbox{I = $200$ pA}), the Mn atoms are still visible in the constant current images and appear as dark contrasts. Because the tunneling current exhibits a exponential dependency on the tip-sample separation, a small change of the tip-sample induces a dramatic change in the tunneling current and thus in the differential conductivity. 
When no voltage is applied between the semiconductor sample and the STM tip, the Fermi levels of the tip and the semiconductor are aligned. We can thus conclude from the spectroscopy measurements that the semiconductor Fermi level is located close to the top of the valence band, which is to be expected for p-type material and further supports our conclusion that Mn acts as an acceptor in InSb. 

The spectroscopy measurements on a Mn atom show an extra current channel at about \mbox{$0.4$ V} which is due to the electrons tunneling from the STM tip into the neutral acceptor state. 
At these tunnel conditions the appearance of the triangular contrast is best observed. The additional current at negative sample voltage measured at the position of a Mn acceptor is due to the influence of the Coulomb potential of the ionized Mn acceptor on the tunnel current.

\begin{figure}[t]
\begin{center}
\includegraphics[width=82.5mm]{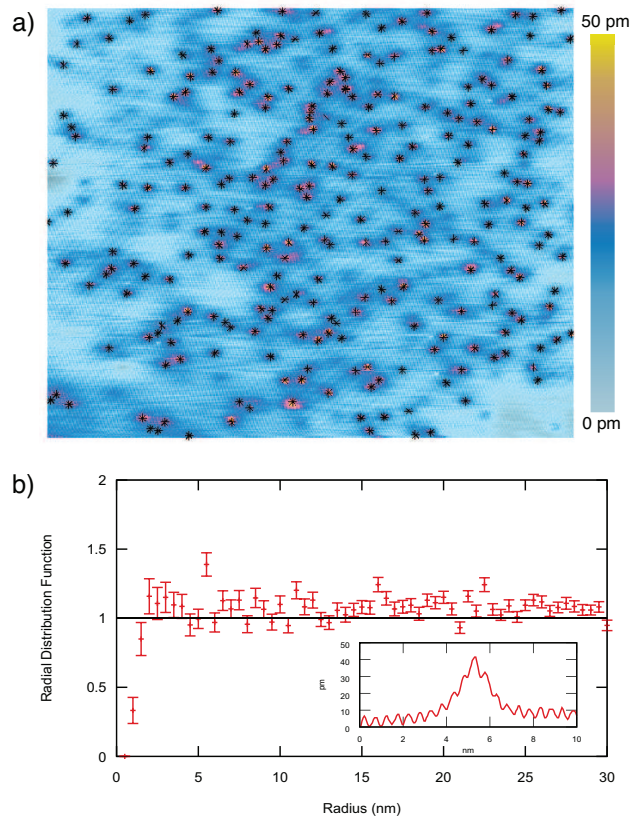}
\caption{(a) Filled state X-STM topography (\mbox{$100$ nm$\times100$ nm}) of ionized Mn acceptors (black stars) near the (110) InSb surface. (\mbox{$V_{sample}$ = $-0.8$ V} and \mbox{I = $200$ pA}) The topographic contrast extends over 50 pm and is represented by a color scale where yellow corresponds with the tip retracted furthest away from the surface. (b) Radial distribution function of the Mn acceptors observed in (a). Inset: Topographic profile line along the [110] direction of a Mn acceptor in InSb observed in \mbox{Fig. \ref{dopant order}(a)} in its ionized charge state.}
\label{dopant order}
\end{center}
\end{figure}

In order to determine the randomness of the spatial distribution of the Mn atoms in the InSb sample we used a filled state image mode (\mbox{$V_{sample}$ = $-0.8$ V} and \mbox{I = $200$ pA}) at which mainly Mn atoms in the outermost surface layer are detected, see \mbox{Fig. \ref{dopant order}(a)}.  At these tunnel conditions the Mn atoms are observed in their ionized state which allows for an optimal spatial resolution of about 3 nm (see inset in \mbox{Fig. \ref{dopant order}(b)}). We used an automated procedure to determine the position of each Mn atom in the image (marked by a black star) and from these dopant positions the radial distribution function was derived, \mbox{Fig. \ref{dopant order}(b)}. The value of the radial distribution function is close to unity for all distances beyond 3 nm, proving that the Mn-distribution is purely random at these length scales. The dip in the radial distribution below 3 nm is due to the X-STM technique which does not allow to separate two Mn atoms that are closer to each other than the resolution limit under these tunnel conditions. If we image the surface at positive voltage, when the Mn atoms are in their neutral state, we can recognize close pairs more easily.\cite{Yakunin2005} However we did not observe any indication that excessive Mn-pairing, as was for instance suggested to occur (In,Mn)As\cite{Soo2004}, is taking place in our (In,Mn)Sb sample.

The image in \mbox{Fig. \ref{dopant order}(a)} shows topographic contrast, i.e. the tip is retracted due to an enhanced tunneling probability, around each ionized Mn atom. This topographic contrast is at maximum 50 pm and extends laterally over about 3 nm. The height contrast we observe in fig. 5 a) is due to a spatial variation in the integrated density-of-states which is available for tunneling from the semiconductor to the tip, i.e. the total density-of-states that is present in the energy range between the top of the valence band and the tip Fermi-level. The contrast varies because the local Coulomb potential around each charged acceptor (the Mn atoms are ionized at the applied tunnel condition) is not constant. This will locally shift the top of the valence band.\cite{Teichmann2008} Because the tip is operating in a constant current mode the height of the tip (the topographic contrast) is adjusted such that the tunnel current remains constant when more states become available for tunneling. In the present case the tip is retracted when the energy interval over which electrons can be extracted from the valence band locally increases due to a variation in the local Coulomb potential. The dark blue regions in the topographic are thus attributed to the presence of the local Coulomb potential from the ionized Mn acceptor. 
It appears that the areas with enhanced contrast, the dark blue regions which correspond with a higher local electrostatic potential, are forming interconnected networks of local potential variations. The holes, donated by the Mn-acceptors, will modify their spatial distribution according to the local potential network and form conductive percolation pathways through the InSb matrix. Along such pathways stronger interactions between the magnetic ions might occur but it is difficult to estimate how this will exactly affect the magnetic behavior. It will also be important to consider that the density of the magnetic moments located on the Mn ions varies. Seemingly conflicting theoretical predictions are made in the literature. For instance in the paper of Sato et al \cite{Sato2004} it is shown, that if percolation is taken into account in the modeling of the magnetic behavior, it will give a much lower Curie temperature as is predicted by a mean-field calculation. However in the theoretical model for DMS by Kennett et al.\cite{kennett2002} of an inhomogeneous DMS it is predicted that an inhomogeneous magnetic material can lead to a higher $T_{C}$ than is found for a material with an ordered lattice of Mn spins.
It is not possible for us to make a conclusive statement based on the X-STM results alone why MOVPE grown InSb shows room-temperature ferromagnetism. It is tempting to use the results of Kennett et al. \cite{kennett2002} but the enhancement they find happens when a disordered system is compared to an idealized system of ordered Mn spins. It remains of high interest to explore further the role of randomness in the distribution of magnetic acceptors, considering both the role of the spatial location of the magnetic moments and the related percolating network of the carriers that can be involved in coupling the magnetic moments.  Furthermore it will be interesting to study how such a  system of magnetic atoms interacts when the carrier concentration is enhanced. Recent cyclotron resonance measurements on similar MOVPE grown InSb samples from the same group indicate that the hole concentration in MOVPE grown samples is higher as in MBE grown samples.\cite{Khodaparast2013} This fact might be important for the room-temperature ferromagnetism observed in MOVPE grown InSb.

In conclusion the incorporation of Mn and its spatial distribution in diluted, magnetic semiconductor (In,Mn)Sb was investigated by X-STM. It was found that Mn atoms incorporates as Mn$^{2+}$ ions and thus acts as an acceptor. The observed contrast of neutral Mn atoms agrees with a small binding energy for the hole bound to this acceptor. The impurities are randomly distributed for pair distances greater than \mbox{$3$ nm} which is the resolution limit of our approach. No indications were found for enhanced pairing or clustering of the Mn on a short length scale or the formation of second phase inclusions. Well-resolved percolation pathways associated with the Mn acceptor are observed in the X-STM images. These pathways might be important for understanding the observed magnetic behavior at high temperatures.

\begin{acknowledgments}
We would like to acknowledge Mu Wang and Bryn Howells from the University of Nottingham for providing support for the SQUID measurements. The research leading to these results has received funding from the European Community's Seventh Framework Programme (PF7/2007-2013) under Grant agreement No. 215368 and from NSF DMR-0804479 and DMR-1305666 .
\end{acknowledgments}

\bibliographystyle{apsrev4-1}
\bibliography{Citations}

%merlin.mbs apsrev4-1.bst 2010-07-25 4.21a (PWD, AO, DPC) hacked
%Control: key (0)
%Control: author (72) initials jnrlst
%Control: editor formatted (1) identically to author
%Control: production of article title (-1) disabled
%Control: page (0) single
%Control: year (1) truncated
%Control: production of eprint (0) enabled
\begin{thebibliography}{27}%
\makeatletter
\providecommand \@ifxundefined [1]{%
 \@ifx{#1\undefined}
}%
\providecommand \@ifnum [1]{%
 \ifnum #1\expandafter \@firstoftwo
 \else \expandafter \@secondoftwo
 \fi
}%
\providecommand \@ifx [1]{%
 \ifx #1\expandafter \@firstoftwo
 \else \expandafter \@secondoftwo
 \fi
}%
\providecommand \natexlab [1]{#1}%
\providecommand \enquote  [1]{``#1''}%
\providecommand \bibnamefont  [1]{#1}%
\providecommand \bibfnamefont [1]{#1}%
\providecommand \citenamefont [1]{#1}%
\providecommand \href@noop [0]{\@secondoftwo}%
\providecommand \href [0]{\begingroup \@sanitize@url \@href}%
\providecommand \@href[1]{\@@startlink{#1}\@@href}%
\providecommand \@@href[1]{\endgroup#1\@@endlink}%
\providecommand \@sanitize@url [0]{\catcode `\\12\catcode `\$12\catcode
  `\&12\catcode `\#12\catcode `\^12\catcode `\_12\catcode `\%12\relax}%
\providecommand \@@startlink[1]{}%
\providecommand \@@endlink[0]{}%
\providecommand \url  [0]{\begingroup\@sanitize@url \@url }%
\providecommand \@url [1]{\endgroup\@href {#1}{\urlprefix }}%
\providecommand \urlprefix  [0]{URL }%
\providecommand \Eprint [0]{\href }%
\providecommand \doibase [0]{http://dx.doi.org/}%
\providecommand \selectlanguage [0]{\@gobble}%
\providecommand \bibinfo  [0]{\@secondoftwo}%
\providecommand \bibfield  [0]{\@secondoftwo}%
\providecommand \translation [1]{[#1]}%
\providecommand \BibitemOpen [0]{}%
\providecommand \bibitemStop [0]{}%
\providecommand \bibitemNoStop [0]{.\EOS\space}%
\providecommand \EOS [0]{\spacefactor3000\relax}%
\providecommand \BibitemShut  [1]{\csname bibitem#1\endcsname}%
\let\auto@bib@innerbib\@empty
%</preamble>
\bibitem [{\citenamefont {Ohno}(1998)}]{Ohno1998}%
  \BibitemOpen
  \bibfield  {author} {\bibinfo {author} {\bibfnamefont {H.}~\bibnamefont
  {Ohno}},\ }\href@noop {} {\bibfield  {journal} {\bibinfo  {journal}
  {Science}\ }\textbf {\bibinfo {volume} {281}},\ \bibinfo {pages} {951}
  (\bibinfo {year} {1998})}\BibitemShut {NoStop}%
\bibitem [{\citenamefont {Ohno}\ \emph {et~al.}(1999)\citenamefont {Ohno},
  \citenamefont {Young}, \citenamefont {Beschoten}, \citenamefont {Matsukura},
  \citenamefont {Ohno},\ and\ \citenamefont {Awschalom}}]{Ohno1999}%
  \BibitemOpen
  \bibfield  {author} {\bibinfo {author} {\bibfnamefont {Y.}~\bibnamefont
  {Ohno}}, \bibinfo {author} {\bibfnamefont {D.~K.}\ \bibnamefont {Young}},
  \bibinfo {author} {\bibfnamefont {B.}~\bibnamefont {Beschoten}}, \bibinfo
  {author} {\bibfnamefont {F.}~\bibnamefont {Matsukura}}, \bibinfo {author}
  {\bibfnamefont {H.}~\bibnamefont {Ohno}}, \ and\ \bibinfo {author}
  {\bibfnamefont {D.~D.}\ \bibnamefont {Awschalom}},\ }\href@noop {} {\bibfield
   {journal} {\bibinfo  {journal} {Nature}\ }\textbf {\bibinfo {volume}
  {402}},\ \bibinfo {pages} {790} (\bibinfo {year} {1999})}\BibitemShut
  {NoStop}%
\bibitem [{\citenamefont {Dietl}\ \emph {et~al.}(2000)\citenamefont {Dietl},
  \citenamefont {Ohno}, \citenamefont {Matsukura}, \citenamefont {Cibert},\
  and\ \citenamefont {Ferrand}}]{Dietl2000}%
  \BibitemOpen
  \bibfield  {author} {\bibinfo {author} {\bibfnamefont {T.}~\bibnamefont
  {Dietl}}, \bibinfo {author} {\bibfnamefont {H.}~\bibnamefont {Ohno}},
  \bibinfo {author} {\bibfnamefont {F.}~\bibnamefont {Matsukura}}, \bibinfo
  {author} {\bibfnamefont {J.}~\bibnamefont {Cibert}}, \ and\ \bibinfo {author}
  {\bibfnamefont {D.}~\bibnamefont {Ferrand}},\ }\href@noop {} {\bibfield
  {journal} {\bibinfo  {journal} {Science}\ }\textbf {\bibinfo {volume}
  {287}},\ \bibinfo {pages} {1019} (\bibinfo {year} {2000})}\BibitemShut
  {NoStop}%
\bibitem [{\citenamefont {Blattner}\ and\ \citenamefont
  {Wessels}(2002)}]{Blattner2002}%
  \BibitemOpen
  \bibfield  {author} {\bibinfo {author} {\bibfnamefont {A.}~\bibnamefont
  {Blattner}}\ and\ \bibinfo {author} {\bibfnamefont {B.}~\bibnamefont
  {Wessels}},\ }in\ \href@noop {} {\emph {\bibinfo {booktitle} {Papers from
  29th Conference on the Physics and Chemistry of Semiconductor Interfaces}}},\
  Vol.~\bibinfo {volume} {20}\ (\bibinfo {year} {2002})\ p.\ \bibinfo {pages}
  {1582}\BibitemShut {NoStop}%
\bibitem [{\citenamefont {Blattner}\ \emph {et~al.}(2001)\citenamefont
  {Blattner}, \citenamefont {Lensch},\ and\ \citenamefont
  {Wessels}}]{Blattner2001}%
  \BibitemOpen
  \bibfield  {author} {\bibinfo {author} {\bibfnamefont {A.}~\bibnamefont
  {Blattner}}, \bibinfo {author} {\bibfnamefont {J.}~\bibnamefont {Lensch}}, \
  and\ \bibinfo {author} {\bibfnamefont {B.}~\bibnamefont {Wessels}},\
  }\href@noop {} {\bibfield  {journal} {\bibinfo  {journal} {Journal of
  Electronic Materials}\ }\textbf {\bibinfo {volume} {30}},\ \bibinfo {pages}
  {1408} (\bibinfo {year} {2001})}\BibitemShut {NoStop}%
\bibitem [{\citenamefont {Parashar}\ \emph {et~al.}(2010)\citenamefont
  {Parashar}, \citenamefont {Rangaraju}, \citenamefont {Lazarov}, \citenamefont
  {Xie},\ and\ \citenamefont {Wessels}}]{Parashar2010}%
  \BibitemOpen
  \bibfield  {author} {\bibinfo {author} {\bibfnamefont {N.~D.}\ \bibnamefont
  {Parashar}}, \bibinfo {author} {\bibfnamefont {N.}~\bibnamefont {Rangaraju}},
  \bibinfo {author} {\bibfnamefont {V.~K.}\ \bibnamefont {Lazarov}}, \bibinfo
  {author} {\bibfnamefont {S.}~\bibnamefont {Xie}}, \ and\ \bibinfo {author}
  {\bibfnamefont {B.~W.}\ \bibnamefont {Wessels}},\ }\href@noop {} {\bibfield
  {journal} {\bibinfo  {journal} {Physical Review B}\ }\textbf {\bibinfo
  {volume} {81}},\ \bibinfo {pages} {115321} (\bibinfo {year}
  {2010})}\BibitemShut {NoStop}%
\bibitem [{\citenamefont {Wessels}(2008)}]{Wessels2008}%
  \BibitemOpen
  \bibfield  {author} {\bibinfo {author} {\bibfnamefont {B.~W.}\ \bibnamefont
  {Wessels}},\ }\href@noop {} {\bibfield  {journal} {\bibinfo  {journal} {New
  Journal Of Physics}\ }\textbf {\bibinfo {volume} {10}},\ \bibinfo {pages}
  {055008} (\bibinfo {year} {2008})}\BibitemShut {NoStop}%
\bibitem [{\citenamefont {Yakunin}\ \emph {et~al.}(2004)\citenamefont
  {Yakunin}, \citenamefont {Silov}, \citenamefont {Koenraad}, \citenamefont
  {Wolter}, \citenamefont {Van~Roy}, \citenamefont {De~Boeck}, \citenamefont
  {Tang},\ and\ \citenamefont {Flatte}}]{Yakunin2004}%
  \BibitemOpen
  \bibfield  {author} {\bibinfo {author} {\bibfnamefont {A.~M.}\ \bibnamefont
  {Yakunin}}, \bibinfo {author} {\bibfnamefont {A.~Y.}\ \bibnamefont {Silov}},
  \bibinfo {author} {\bibfnamefont {P.~M.}\ \bibnamefont {Koenraad}}, \bibinfo
  {author} {\bibfnamefont {J.~H.}\ \bibnamefont {Wolter}}, \bibinfo {author}
  {\bibfnamefont {W.}~\bibnamefont {Van~Roy}}, \bibinfo {author} {\bibfnamefont
  {J.}~\bibnamefont {De~Boeck}}, \bibinfo {author} {\bibfnamefont {J.~M.}\
  \bibnamefont {Tang}}, \ and\ \bibinfo {author} {\bibfnamefont {M.~E.}\
  \bibnamefont {Flatte}},\ }\href@noop {} {\bibfield  {journal} {\bibinfo
  {journal} {Physical Review Letters}\ }\textbf {\bibinfo {volume} {92}},\
  \bibinfo {pages} {216806} (\bibinfo {year} {2004})}\BibitemShut {NoStop}%
\bibitem [{\citenamefont {Yakunin}\ \emph {et~al.}(2005)\citenamefont
  {Yakunin}, \citenamefont {Silov}, \citenamefont {Koenraad}, \citenamefont
  {Tang}, \citenamefont {Flatte}, \citenamefont {Van~Roy}, \citenamefont
  {De~Boeck},\ and\ \citenamefont {Wolter}}]{Yakunin2005}%
  \BibitemOpen
  \bibfield  {author} {\bibinfo {author} {\bibfnamefont {A.~M.}\ \bibnamefont
  {Yakunin}}, \bibinfo {author} {\bibfnamefont {A.~Y.}\ \bibnamefont {Silov}},
  \bibinfo {author} {\bibfnamefont {P.~M.}\ \bibnamefont {Koenraad}}, \bibinfo
  {author} {\bibfnamefont {J.~M.}\ \bibnamefont {Tang}}, \bibinfo {author}
  {\bibfnamefont {M.~E.}\ \bibnamefont {Flatte}}, \bibinfo {author}
  {\bibfnamefont {W.}~\bibnamefont {Van~Roy}}, \bibinfo {author} {\bibfnamefont
  {J.}~\bibnamefont {De~Boeck}}, \ and\ \bibinfo {author} {\bibfnamefont
  {J.~H.}\ \bibnamefont {Wolter}},\ }\href@noop {} {\bibfield  {journal}
  {\bibinfo  {journal} {Physical Review Letters}\ }\textbf {\bibinfo {volume}
  {95}},\ \bibinfo {pages} {256402} (\bibinfo {year} {2005})}\BibitemShut
  {NoStop}%
\bibitem [{\citenamefont {Sato}\ \emph {et~al.}(2004)\citenamefont {Sato},
  \citenamefont {Schweika}, \citenamefont {Dederichs},\ and\ \citenamefont
  {Katayama-Yoshida}}]{Sato2004}%
  \BibitemOpen
  \bibfield  {author} {\bibinfo {author} {\bibfnamefont {K.}~\bibnamefont
  {Sato}}, \bibinfo {author} {\bibfnamefont {W.}~\bibnamefont {Schweika}},
  \bibinfo {author} {\bibfnamefont {P.~H.}\ \bibnamefont {Dederichs}}, \ and\
  \bibinfo {author} {\bibfnamefont {H.}~\bibnamefont {Katayama-Yoshida}},\
  }\href@noop {} {\bibfield  {journal} {\bibinfo  {journal} {Physical Review
  B}\ }\textbf {\bibinfo {volume} {70}},\ \bibinfo {pages} {201202(R)}
  (\bibinfo {year} {2004})}\BibitemShut {NoStop}%
\bibitem [{\citenamefont {Kennett}\ \emph {et~al.}(2004)\citenamefont
  {Kennett}, \citenamefont {Berciu},\ and\ \citenamefont
  {Bhatt}}]{Kennett2004}%
  \BibitemOpen
  \bibfield  {author} {\bibinfo {author} {\bibfnamefont {M.~P.}\ \bibnamefont
  {Kennett}}, \bibinfo {author} {\bibfnamefont {M.}~\bibnamefont {Berciu}}, \
  and\ \bibinfo {author} {\bibfnamefont {R.~N.}\ \bibnamefont {Bhatt}},\
  }\href@noop {} {\bibfield  {journal} {\bibinfo  {journal} {Journal of
  Magnetism and Magnetic Materials}\ }\textbf {\bibinfo {volume} {272}},\
  \bibinfo {pages} {1993} (\bibinfo {year} {2004})}\BibitemShut {NoStop}%
\bibitem [{\citenamefont {Bouzerar}\ \emph {et~al.}(2004)\citenamefont
  {Bouzerar}, \citenamefont {Ziman},\ and\ \citenamefont
  {Kudrnovsk{\'y}}}]{Bouzerar2004}%
  \BibitemOpen
  \bibfield  {author} {\bibinfo {author} {\bibfnamefont {G.}~\bibnamefont
  {Bouzerar}}, \bibinfo {author} {\bibfnamefont {T.}~\bibnamefont {Ziman}}, \
  and\ \bibinfo {author} {\bibfnamefont {J.}~\bibnamefont {Kudrnovsk{\'y}}},\
  }\href@noop {} {\bibfield  {journal} {\bibinfo  {journal} {Applied Physics
  Letters}\ }\textbf {\bibinfo {volume} {85}},\ \bibinfo {pages} {4941}
  (\bibinfo {year} {2004})}\BibitemShut {NoStop}%
\bibitem [{\citenamefont {Soo}\ \emph {et~al.}(2004)\citenamefont {Soo},
  \citenamefont {Kim}, \citenamefont {Kao}, \citenamefont {Blattner},
  \citenamefont {Wessels}, \citenamefont {Khalid}, \citenamefont
  {Sanchez~Hanke},\ and\ \citenamefont {Kao}}]{Soo2004}%
  \BibitemOpen
  \bibfield  {author} {\bibinfo {author} {\bibfnamefont {Y.~L.}\ \bibnamefont
  {Soo}}, \bibinfo {author} {\bibfnamefont {S.}~\bibnamefont {Kim}}, \bibinfo
  {author} {\bibfnamefont {Y.~H.}\ \bibnamefont {Kao}}, \bibinfo {author}
  {\bibfnamefont {A.~J.}\ \bibnamefont {Blattner}}, \bibinfo {author}
  {\bibfnamefont {B.~W.}\ \bibnamefont {Wessels}}, \bibinfo {author}
  {\bibfnamefont {S.}~\bibnamefont {Khalid}}, \bibinfo {author} {\bibfnamefont
  {C.}~\bibnamefont {Sanchez~Hanke}}, \ and\ \bibinfo {author} {\bibfnamefont
  {C.~C.}\ \bibnamefont {Kao}},\ }\href@noop {} {\bibfield  {journal} {\bibinfo
   {journal} {Applied Physics Letters}\ }\textbf {\bibinfo {volume} {84}},\
  \bibinfo {pages} {481} (\bibinfo {year} {2004})}\BibitemShut {NoStop}%
\bibitem [{\citenamefont {Kitchen}\ \emph {et~al.}(2006)\citenamefont
  {Kitchen}, \citenamefont {Richardella}, \citenamefont {Tang}, \citenamefont
  {Flatte},\ and\ \citenamefont {Yazdani}}]{Kitchen2006}%
  \BibitemOpen
  \bibfield  {author} {\bibinfo {author} {\bibfnamefont {D.}~\bibnamefont
  {Kitchen}}, \bibinfo {author} {\bibfnamefont {A.}~\bibnamefont
  {Richardella}}, \bibinfo {author} {\bibfnamefont {J.-M.}\ \bibnamefont
  {Tang}}, \bibinfo {author} {\bibfnamefont {M.~E.}\ \bibnamefont {Flatte}}, \
  and\ \bibinfo {author} {\bibfnamefont {A.}~\bibnamefont {Yazdani}},\
  }\href@noop {} {\bibfield  {journal} {\bibinfo  {journal} {Nature}\ }\textbf
  {\bibinfo {volume} {442}},\ \bibinfo {pages} {436} (\bibinfo {year}
  {2006})}\BibitemShut {NoStop}%
\bibitem [{\citenamefont {Richardella}\ \emph {et~al.}(2010)\citenamefont
  {Richardella}, \citenamefont {Roushan}, \citenamefont {Mack}, \citenamefont
  {Zhou}, \citenamefont {Huse}, \citenamefont {Awschalom},\ and\ \citenamefont
  {Yazdani}}]{Richardella2010}%
  \BibitemOpen
  \bibfield  {author} {\bibinfo {author} {\bibfnamefont {A.}~\bibnamefont
  {Richardella}}, \bibinfo {author} {\bibfnamefont {P.}~\bibnamefont
  {Roushan}}, \bibinfo {author} {\bibfnamefont {S.}~\bibnamefont {Mack}},
  \bibinfo {author} {\bibfnamefont {B.}~\bibnamefont {Zhou}}, \bibinfo {author}
  {\bibfnamefont {D.}~\bibnamefont {Huse}}, \bibinfo {author} {\bibfnamefont
  {D.}~\bibnamefont {Awschalom}}, \ and\ \bibinfo {author} {\bibfnamefont
  {A.}~\bibnamefont {Yazdani}},\ }\href@noop {} {\bibfield  {journal} {\bibinfo
   {journal} {Science}\ }\textbf {\bibinfo {volume} {327}},\ \bibinfo {pages}
  {665} (\bibinfo {year} {2010})}\BibitemShut {NoStop}%
\bibitem [{\citenamefont {Bozkurt}\ \emph {et~al.}(2010)\citenamefont
  {Bozkurt}, \citenamefont {Grant}, \citenamefont {Ulloa}, \citenamefont
  {Campion}, \citenamefont {Foxon}, \citenamefont {Marega}, \citenamefont
  {Salamo},\ and\ \citenamefont {Koenraad}}]{Bozkurt2010}%
  \BibitemOpen
  \bibfield  {author} {\bibinfo {author} {\bibfnamefont {M.}~\bibnamefont
  {Bozkurt}}, \bibinfo {author} {\bibfnamefont {A.}~\bibnamefont {Grant}},
  \bibinfo {author} {\bibfnamefont {J.~M.}\ \bibnamefont {Ulloa}}, \bibinfo
  {author} {\bibfnamefont {R.~P.}\ \bibnamefont {Campion}}, \bibinfo {author}
  {\bibfnamefont {C.~T.}\ \bibnamefont {Foxon}}, \bibinfo {author}
  {\bibfnamefont {E.}~\bibnamefont {Marega}}, \bibinfo {author} {\bibfnamefont
  {G.~J.}\ \bibnamefont {Salamo}}, \ and\ \bibinfo {author} {\bibfnamefont
  {P.~M.}\ \bibnamefont {Koenraad}},\ }\href@noop {} {\bibfield  {journal}
  {\bibinfo  {journal} {Applied Physics Letters}\ }\textbf {\bibinfo {volume}
  {96}},\ \bibinfo {pages} {042108} (\bibinfo {year} {2010})}\BibitemShut
  {NoStop}%
\bibitem [{\citenamefont {Parashar}\ \emph {et~al.}(2009)\citenamefont
  {Parashar}, \citenamefont {Keavney},\ and\ \citenamefont
  {Wessels}}]{Parashar2009}%
  \BibitemOpen
  \bibfield  {author} {\bibinfo {author} {\bibfnamefont {N.~D.}\ \bibnamefont
  {Parashar}}, \bibinfo {author} {\bibfnamefont {D.~J.}\ \bibnamefont
  {Keavney}}, \ and\ \bibinfo {author} {\bibfnamefont {B.~W.}\ \bibnamefont
  {Wessels}},\ }\href@noop {} {\bibfield  {journal} {\bibinfo  {journal}
  {Applied Physics Letters}\ }\textbf {\bibinfo {volume} {95}},\  (\bibinfo
  {year} {2009})}\BibitemShut {NoStop}%
\bibitem [{\citenamefont {Celebi}\ \emph {et~al.}(2010)\citenamefont {Celebi},
  \citenamefont {Garleff}, \citenamefont {Silov}, \citenamefont {Yakunin},
  \citenamefont {Koenraad}, \citenamefont {Van~Roy}, \citenamefont {Tang},\
  and\ \citenamefont {Flatte}}]{Celebi2010}%
  \BibitemOpen
  \bibfield  {author} {\bibinfo {author} {\bibfnamefont {C.}~\bibnamefont
  {Celebi}}, \bibinfo {author} {\bibfnamefont {J.~K.}\ \bibnamefont {Garleff}},
  \bibinfo {author} {\bibfnamefont {A.~Y.}\ \bibnamefont {Silov}}, \bibinfo
  {author} {\bibfnamefont {A.~M.}\ \bibnamefont {Yakunin}}, \bibinfo {author}
  {\bibfnamefont {P.~M.}\ \bibnamefont {Koenraad}}, \bibinfo {author}
  {\bibfnamefont {W.}~\bibnamefont {Van~Roy}}, \bibinfo {author} {\bibfnamefont
  {J.-M.}\ \bibnamefont {Tang}}, \ and\ \bibinfo {author} {\bibfnamefont
  {M.~E.}\ \bibnamefont {Flatte}},\ }\href@noop {} {\bibfield  {journal}
  {\bibinfo  {journal} {Physical Review Letters}\ }\textbf {\bibinfo {volume}
  {104}},\ \bibinfo {pages} {086404} (\bibinfo {year} {2010})}\BibitemShut
  {NoStop}%
\bibitem [{\citenamefont {Zheng}\ \emph {et~al.}(1994)\citenamefont {Zheng},
  \citenamefont {Salmeron},\ and\ \citenamefont {Weber}}]{Zheng1994}%
  \BibitemOpen
  \bibfield  {author} {\bibinfo {author} {\bibfnamefont {Z.~F.}\ \bibnamefont
  {Zheng}}, \bibinfo {author} {\bibfnamefont {M.~B.}\ \bibnamefont {Salmeron}},
  \ and\ \bibinfo {author} {\bibfnamefont {E.~R.}\ \bibnamefont {Weber}},\
  }\href@noop {} {\bibfield  {journal} {\bibinfo  {journal} {Applied Physics
  Letters}\ }\textbf {\bibinfo {volume} {64}},\ \bibinfo {pages} {1836}
  (\bibinfo {year} {1994})}\BibitemShut {NoStop}%
\bibitem [{\citenamefont {Marczinowski}\ \emph {et~al.}(2008)\citenamefont
  {Marczinowski}, \citenamefont {Wiebe}, \citenamefont {Meier}, \citenamefont
  {Hashimoto},\ and\ \citenamefont {Wiesendanger}}]{Marczinowski2008}%
  \BibitemOpen
  \bibfield  {author} {\bibinfo {author} {\bibfnamefont {F.}~\bibnamefont
  {Marczinowski}}, \bibinfo {author} {\bibfnamefont {J.}~\bibnamefont {Wiebe}},
  \bibinfo {author} {\bibfnamefont {F.}~\bibnamefont {Meier}}, \bibinfo
  {author} {\bibfnamefont {K.}~\bibnamefont {Hashimoto}}, \ and\ \bibinfo
  {author} {\bibfnamefont {R.}~\bibnamefont {Wiesendanger}},\ }\href@noop {}
  {\bibfield  {journal} {\bibinfo  {journal} {Physical Review B}\ }\textbf
  {\bibinfo {volume} {77}},\ \bibinfo {pages} {115318} (\bibinfo {year}
  {2008})}\BibitemShut {NoStop}%
\bibitem [{\citenamefont {Obukhov}\ \emph {et~al.}(1991)\citenamefont
  {Obukhov}, \citenamefont {Neganov}, \citenamefont {Kiselev}, \citenamefont
  {Chemikov}, \citenamefont {Vekshina}, \citenamefont {Pepik},\ and\
  \citenamefont {Popkov}}]{Obukhov1991}%
  \BibitemOpen
  \bibfield  {author} {\bibinfo {author} {\bibfnamefont {S.~A.}\ \bibnamefont
  {Obukhov}}, \bibinfo {author} {\bibfnamefont {B.~S.}\ \bibnamefont
  {Neganov}}, \bibinfo {author} {\bibfnamefont {Y.~F.}\ \bibnamefont
  {Kiselev}}, \bibinfo {author} {\bibfnamefont {A.~N.}\ \bibnamefont
  {Chemikov}}, \bibinfo {author} {\bibfnamefont {V.~S.}\ \bibnamefont
  {Vekshina}}, \bibinfo {author} {\bibfnamefont {N.~I.}\ \bibnamefont {Pepik}},
  \ and\ \bibinfo {author} {\bibfnamefont {A.~N.}\ \bibnamefont {Popkov}},\
  }\href@noop {} {\bibfield  {journal} {\bibinfo  {journal} {Cryogenics}\
  }\textbf {\bibinfo {volume} {31}},\ \bibinfo {pages} {874} (\bibinfo {year}
  {1991})}\BibitemShut {NoStop}%
\bibitem [{\citenamefont {Obukhov}(2005)}]{Obukhov2005}%
  \BibitemOpen
  \bibfield  {author} {\bibinfo {author} {\bibfnamefont {S.~A.}\ \bibnamefont
  {Obukhov}},\ }\href@noop {} {\bibfield  {journal} {\bibinfo  {journal} {Phys.
  Stat. Sol. (b)}\ }\textbf {\bibinfo {volume} {242}},\ \bibinfo {pages} {1298}
  (\bibinfo {year} {2005})}\BibitemShut {NoStop}%
\bibitem [{\citenamefont {Teubert}\ \emph {et~al.}(2009)\citenamefont
  {Teubert}, \citenamefont {Obukhov}, \citenamefont {Klar},\ and\ \citenamefont
  {Heimbrodt}}]{Teubert2009}%
  \BibitemOpen
  \bibfield  {author} {\bibinfo {author} {\bibfnamefont {J.}~\bibnamefont
  {Teubert}}, \bibinfo {author} {\bibfnamefont {S.~A.}\ \bibnamefont
  {Obukhov}}, \bibinfo {author} {\bibfnamefont {P.~J.}\ \bibnamefont {Klar}}, \
  and\ \bibinfo {author} {\bibfnamefont {W.}~\bibnamefont {Heimbrodt}},\
  }\href@noop {} {\bibfield  {journal} {\bibinfo  {journal} {Physical Review
  Letters}\ }\textbf {\bibinfo {volume} {102}},\ \bibinfo {pages} {046404}
  (\bibinfo {year} {2009})}\BibitemShut {NoStop}%
\bibitem [{\citenamefont {Teichmann}\ \emph {et~al.}(2008)\citenamefont
  {Teichmann}, \citenamefont {Wenderoth}, \citenamefont {Loth}, \citenamefont
  {Ulbrich}, \citenamefont {Garleff}, \citenamefont {Wijnheijmer},\ and\
  \citenamefont {Koenraad}}]{Teichmann2008}%
  \BibitemOpen
  \bibfield  {author} {\bibinfo {author} {\bibfnamefont {K.}~\bibnamefont
  {Teichmann}}, \bibinfo {author} {\bibfnamefont {M.}~\bibnamefont
  {Wenderoth}}, \bibinfo {author} {\bibfnamefont {S.}~\bibnamefont {Loth}},
  \bibinfo {author} {\bibfnamefont {R.~G.}\ \bibnamefont {Ulbrich}}, \bibinfo
  {author} {\bibfnamefont {J.~K.}\ \bibnamefont {Garleff}}, \bibinfo {author}
  {\bibfnamefont {A.~P.}\ \bibnamefont {Wijnheijmer}}, \ and\ \bibinfo {author}
  {\bibfnamefont {P.~M.}\ \bibnamefont {Koenraad}},\ }\href@noop {} {\bibfield
  {journal} {\bibinfo  {journal} {Physical Review Letters}\ }\textbf {\bibinfo
  {volume} {101}},\ \bibinfo {pages} {076103} (\bibinfo {year}
  {2008})}\BibitemShut {NoStop}%
\bibitem [{\citenamefont {Garleff}\ \emph {et~al.}(2008)\citenamefont
  {Garleff}, \citenamefont {Celebi}, \citenamefont {Van~Roy}, \citenamefont
  {Tang}, \citenamefont {Flatte},\ and\ \citenamefont
  {Koenraad}}]{Garleff2008}%
  \BibitemOpen
  \bibfield  {author} {\bibinfo {author} {\bibfnamefont {J.~K.}\ \bibnamefont
  {Garleff}}, \bibinfo {author} {\bibfnamefont {C.}~\bibnamefont {Celebi}},
  \bibinfo {author} {\bibfnamefont {W.}~\bibnamefont {Van~Roy}}, \bibinfo
  {author} {\bibfnamefont {J.~M.}\ \bibnamefont {Tang}}, \bibinfo {author}
  {\bibfnamefont {M.~E.}\ \bibnamefont {Flatte}}, \ and\ \bibinfo {author}
  {\bibfnamefont {P.~M.}\ \bibnamefont {Koenraad}},\ }\href@noop {} {\bibfield
  {journal} {\bibinfo  {journal} {Physical Review B}\ }\textbf {\bibinfo
  {volume} {78}},\ \bibinfo {pages} {075313} (\bibinfo {year}
  {2008})}\BibitemShut {NoStop}%
\bibitem [{\citenamefont {Kennett}\ \emph {et~al.}(2002)\citenamefont
  {Kennett}, \citenamefont {Berciu},\ and\ \citenamefont
  {Bhatt}}]{kennett2002}%
  \BibitemOpen
  \bibfield  {author} {\bibinfo {author} {\bibfnamefont {M.~P.}\ \bibnamefont
  {Kennett}}, \bibinfo {author} {\bibfnamefont {M.}~\bibnamefont {Berciu}}, \
  and\ \bibinfo {author} {\bibfnamefont {R.~N.}\ \bibnamefont {Bhatt}},\
  }\href@noop {} {\bibfield  {journal} {\bibinfo  {journal} {Physical Review
  B}\ }\textbf {\bibinfo {volume} {65}},\ \bibinfo {pages} {115308} (\bibinfo
  {year} {2002})}\BibitemShut {NoStop}%
\bibitem [{\citenamefont {Khodaparast}\ \emph {et~al.}(2013)\citenamefont
  {Khodaparast}, \citenamefont {Matsuda}, \citenamefont {Saha}, \citenamefont
  {Sanders}, \citenamefont {Stanton}, \citenamefont {Saito}, \citenamefont
  {Takeyama}, \citenamefont {Merritt}, \citenamefont {Feeser}, \citenamefont
  {Wessels}, \citenamefont {Liu},\ and\ \citenamefont
  {Furdyna}}]{Khodaparast2013}%
  \BibitemOpen
  \bibfield  {author} {\bibinfo {author} {\bibfnamefont {G.~A.}\ \bibnamefont
  {Khodaparast}}, \bibinfo {author} {\bibfnamefont {Y.~H.}\ \bibnamefont
  {Matsuda}}, \bibinfo {author} {\bibfnamefont {D.}~\bibnamefont {Saha}},
  \bibinfo {author} {\bibfnamefont {G.~D.}\ \bibnamefont {Sanders}}, \bibinfo
  {author} {\bibfnamefont {C.~J.}\ \bibnamefont {Stanton}}, \bibinfo {author}
  {\bibfnamefont {H.}~\bibnamefont {Saito}}, \bibinfo {author} {\bibfnamefont
  {S.}~\bibnamefont {Takeyama}}, \bibinfo {author} {\bibfnamefont {T.~R.}\
  \bibnamefont {Merritt}}, \bibinfo {author} {\bibfnamefont {C.}~\bibnamefont
  {Feeser}}, \bibinfo {author} {\bibfnamefont {B.~W.}\ \bibnamefont {Wessels}},
  \bibinfo {author} {\bibfnamefont {X.}~\bibnamefont {Liu}}, \ and\ \bibinfo
  {author} {\bibfnamefont {J.}~\bibnamefont {Furdyna}},\ }\href@noop {}
  {\bibfield  {journal} {\bibinfo  {journal} {Physical Review B}\ }\textbf
  {\bibinfo {volume} {88}},\ \bibinfo {pages} {235204} (\bibinfo {year}
  {2013})}\BibitemShut {NoStop}%
\end{thebibliography}%

\end{document}